\documentclass[10pt, letterpaper, twocolumn]{article}
\usepackage{ol}
\usepackage{amsmath}
\usepackage{graphicx}
\usepackage{psfrag}

\begin{document}

\twocolumn[
\title{Detuned Twin-Signal-Recycling for ultra-high precision interferometers}

\author{Andr\'{e} Th\"uring, R. Schnabel, H. L\"uck, and K. Danzmann}

\address{Institut f\"ur Gravitationsphysik, Leibniz Universtit\"at Hannover and Max-Planck-Institut f\"ur Gravitationsphysik (Albert-Einstein Institut), Callinstra\ss e 38, 30167 Hannover, Germany}

\begin{abstract}
We propose a new interferometer technique for high precision phase measurements such as those  in
gravitational wave detection. The technique utilizes a pair of optically coupled resonators that provides
identical resonance conditions for the upper as well the lower phase modulation signal sidebands.  This
symmetry significantly reduces the noise spectral density in a wide frequency band compared with single
sideband recycling topologies of current and planned gravitational wave detectors. Furthermore the
application of squeezed states of light becomes less demanding.
\end{abstract}

]

Current interferometric gravitational wave (GW) detectors achieve a phase noise spectral density of as low as  $10^{-11}$ rad$/\sqrt{\mbox{Hz}}$ between 100\,Hz and 1000\,Hz which corresponds to a strain noise spectral density of $10^{-23}$ $1/\sqrt{\mbox{Hz}}$.
The basic topology of those detectors is that of a Michelson interferometer, with km-scale arms. However, to reach such high sensitivities, optical cavities play an essential role.
Arm cavities and power recycling cavities which are tuned to resonance for the laser carrier light, are used in the current GW detectors LIGO~\cite{ligo}, TAMA~\cite{tama} and VIRGO~\cite{virgo}, to increase the circulating light powers. GEO\,600~\cite{geo} uses carrier detuned (single sideband) signal recycling (SR)~\cite{HSMSWWSRD98} that turns the interferometer into a resonant detector. In this scheme the upper (or lower) signal sideband is resonantly enhanced and therefore the sensitivity improved, although the lower sideband (upper, respectively) is suppressed. A similar technique is planned to be used in Advanced\,LIGO~\cite{advligo}.
All the techniques mentioned improve the signal to quantum noise ratio of the detector compared with a simple Michelson interferometer.
Injection of squeezed states is another technique aiming at quantum noise reduction in interferometers
\cite{Cav81}. Current progress in generation and control of squeezed states of light has shown that this
technique can be applied in future detectors, see Ref.~[8] 
and references given therein. Full compatibility of squeezed field injection and detuned SR has been theoretically shown in Harms \emph{et al.} considering homodyne detection~\cite{HCCFVDS03}. Here, `full compatibility' refers to frequency dependent squeezing of quantum  radiation pressure noise and shot noise, providing a broadband sensitivity improvement. Quite generally, two or more low loss, narrow linewidth and therefore long baseline optical filter cavities are required to prepare the squeezed states in an optimum way to achieve this goal, as first proposed in Ref.~[10]. 
However, even in the purely shot noise limited regime, which is typically realized at sideband frequencies above 100\,Hz, still a single filter cavity needs to be applied to prepare the squeezed states~\cite{SHSD04}.

\begin{figure}[ht!]
\includegraphics[scale=0.41]{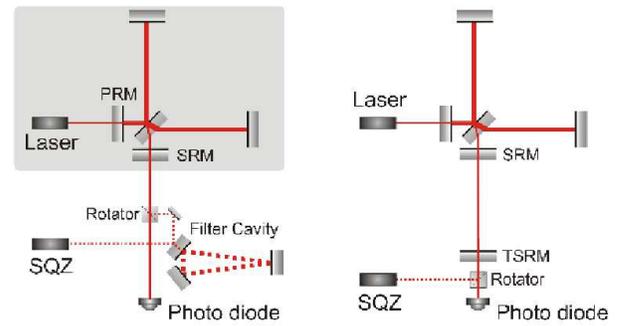}
\caption{\label{schema}(Color online) Top left (shaded): Topology of the current gravitational wave detector GEO\,600. The mirror PRM in the laser input port realizes so-called power-recycling. The signal-recycling mirror (SRM) in the output port establishes a carrier light detuned single-sideband signal recycling cavity.
Bottom left: Extension for a broadband shot-noise reduction utilizing squeezed states.
Right: Topology, proposed here. Two optically coupled cavities are formed with the help of an additional mirror TSRM. Their resonance doublet enables detuned twin-signal-recycling resulting in lower shot noise. Squeezed states can be used without additional filter cavity.}
\end{figure}

In this Letter we propose a new technique for squeezed state enhanced interferometers, namely detuned
\emph{twin-signal-recycling} (TSR). This technique achieves an improved signal to shot noise ratio compared
with the detuned single-sideband signal recycling (SR) technique over a wide part of the detection band,
because upper and lower signal sideband are resonantly enhanced simultaneously~\cite{MAD}. We show that with
this technique squeezed states can be used to reduce the interferometer's shot noise without the need for an
additional filter cavity.

The shaded upper left part of Fig.~\ref{schema} shows a Michelson interferometer topology with the techniques of power- and signal-recycling as used in the GEO\,600 detector. Such an interferometer is operated close to a dark fringe which is a requirement for recycling techniques. To form a detuned signal-recycling cavity, mirror SRM is positioned such that not carrier laser light but signal light is resonating. The lower left part of Fig.~\ref{schema} shows the extension to employ squeezed states of light. A source of a broadband squeezed field (SQZ), a long baseline filter cavity and a Faraday rotator for coupling to the signal modes is required for a broadband sensitivity improvement. A table-top prototype of such an interferometer has recently been demonstrated~\cite {VCHFDS05}.

The right part of Fig.~\ref{schema} shows the topology proposed here. Starting from the standard signal recycling topology, an additional mirror (TSRM) is placed in the signal output port of the interferometer.
This mirror forms a new long-baseline cavity that is optically coupled with the initial signal recycling cavity.
\begin{figure}[h!]
\includegraphics[scale=0.68]{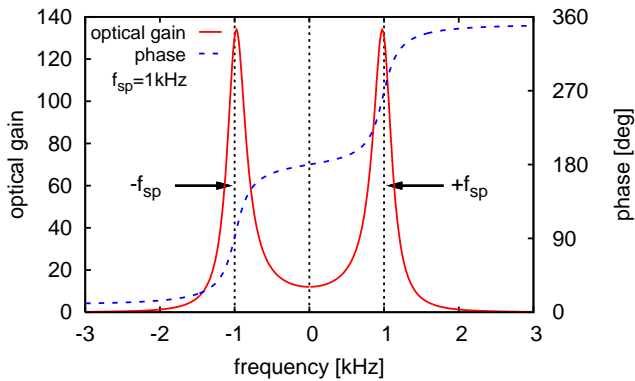}
\caption{\label{doublet}(Color online) Resonance doublet of two optically coupled Fabry-Perot resonators. The
frequency splitting $f_\textrm{sp}=\pm 1\,\textrm{kHz}$ is related to the carrier frequency $f_0$, here set
to zero.}
\end{figure}
Similar to two coupled mechanical oscillators, two coupled optical cavities feature a resonance doublet which
has been experimentally observed, see for example Ref.~[14]. 
The frequency splitting
$\omega_{\text{sp}}$ of this resonance doublet is determined by the coupling of the two resonators, i.e. the
transmission of the center mirror (SRM). The transmission $T_{\text{c}}$ of the mirror SRM corresponding to
this coupling can be obtained from solving the intersection~\cite{ThuringPRE72}
\begin{equation}
-\frac{1}{2}\arg\left[\rho_{23}\left(\frac{\omega_{\text{sp}}L_{\text{SR2}}}{c}\right)\right] =\frac{L_{\text{SR1}}}{L_{\text{SR2}}}\frac{\omega_{\text{sp}}L_{\text{SR2}}}{c}\label{intersec},
\end{equation}
where $L_{\text{SR1}}$ is the length of the resonator in the output port (formed by the mirrors SRM and
TSRM),  and $L_{\text{SR2}}$ is the length of the resonator formed by the SRM and the interferometers' end
mirrors. The term $\arg\left(\rho_{23}\right)$ is the frequency dependent phase of the light field reflected
from this resonator determining the resonance condition of the three-mirror cavity. Notice that the
macroscopic lengths determine the free spectral ranges of the single resonators. The right hand side
of Eq.\,(\ref{intersec}) describes the dispersion in the two resonators, we refer to Ref.~[15] 
 for further details. If lengths $L_{\text{SR1}}$ and $L_{\text{SR2}}$ are assumed to be equal the solution of
Eq.\,(\ref{intersec}) is given by
\begin{equation}
T_{\text{c}}=1-\frac{4\,\cos^2\left(2\frac{\omega_{\text{sp}}L_{\text{SR1}}}{c}\right)\rho_{\text{end}}^2}{\left(1+\rho_{\text{end}}^2\right)^2}\label{tc}.
\end{equation}
For ideal interferometer end mirrors with reflectivity $\rho_{\text{end}}=1$, Eq.\,(\ref{tc}) further reduces to
\begin{equation}
T_{\text{c}}=1-\cos^2\left(2\frac{\omega_{\text{sp}}L_{\text{SR1}}}{c}\right).
\end{equation}
Thus, the transmission $T_{\text{c}}$ of the  mirror SRM has to be chosen with respect to the required
frequency splitting $\omega_{\text{sp}}=2\pi f_{\text{sp}}$. Then the resonances' bandwidth of the doublet can be determined by the reflectivity of the mirror TSRM.
In this Letter we propose equal lengths $L_{\text{SR1}}$  and $L_{\text{SR2}}$ and in the following a frequency splitting $f_{\text{sp}}=1\,{\text{kHz}}$. The reflectivity of TSRM is chosen to give the same peak sensitivity as in the SR case. Fig.~\ref{doublet} shows the resonance doublet for $L_{\text{SR1}} = L_{\text{SR2}}=1200$\,m, $\rho_{\text{MI}}^2=0.99995$ and $\rho_{\text{TSRM}}^2 = 0.963$. The resonance doublet is symmetric around the carrier frequency $f_0$ in magnitude and phase, respectively. Thus, if the three-mirror coupled cavity is tuned to $f_0$, upper and lower signal sidebands have identical resonance conditions.

\begin{figure}[h!]
\psfrag{NSD [units]}{\fontsize{7}{7}{$\text{NSD}$ [$1/\sqrt{\text{Hz}}$]}}
\includegraphics[scale=0.66]{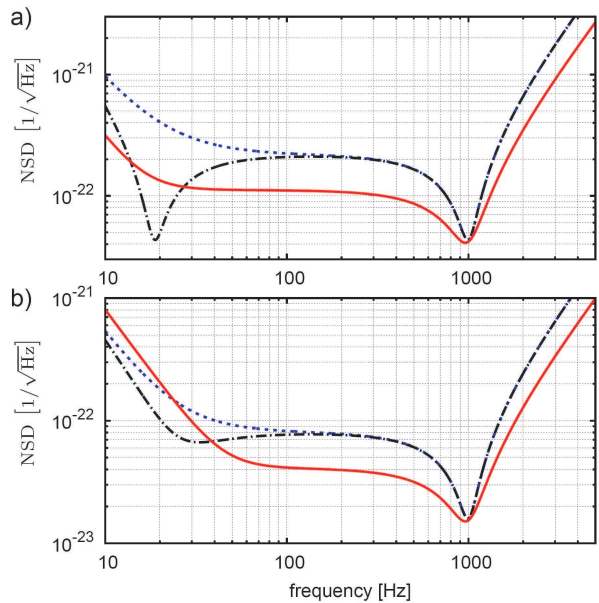}
\caption{\label{sqzinput}(Color online) Comparison of linear noise spectral densities (NSD) of
twin-signal-recycling (TSR) (solid lines) with detuned single sideband signal-recycling (SR) a) without
squeezed input and b) with squeezed input ($r=1$), respectively. The dashed dotted curves represent SR of the
upper sideband; the dotted curves represent SR of the lower sideband. In both cases the TSR topology
yields a lower noise spectral density, by up to a factor of two, at sideband frequencies above 40\,Hz. For a detailed description refer to the text.}
\end{figure}

Fig.~\ref{sqzinput} shows the quantum noise spectral densities of detuned TSR (solid
lines) in comparison with detuned SR using homodyne detection at the signal port. All traces
were calculated using technical design parameters of the GEO\,600 detector as used in Ref.~[9].
In the case of detuned SR the spectral noise density is different for detection of
amplitude and phase quadrature. Here, we consider detection of the amplitude quadrature which provides low noise spectral densities for frequencies below the optical resonance, here at 1\,kHz.

Fig.~\ref{sqzinput}\,a compares SR and TSR without squeezed field input. The dotted (dotted-dashed) curve corresponds to SR with resonating lower (upper) signal sideband at about 1\,kHz. For signal frequencies above 40\,Hz the TSR technique proposed here provides an improved noise spectral density. This frequency range corresponds to the shot-noise dominated regime.
For lower frequencies, quantum radiation pressure noise dominates shot noise and the detuned SR topology with resonating lower signal sideband shows a resonance dip at slightly below 20\,Hz. This is the so-called \emph{optical spring} resonance~\cite{BuonannoPRD65}. Such a resonance does not appear in the TSR topology proposed here, because upper and lower sidebands have detunings symmetrically arranged around the carrier frequency $\omega_0$.
Note that in the case of TSR, phase modulation as introduced by gravitational waves only appears in the carrier lights phase quadrature as it is expected from a simple Michelson interferometer. This contrasts detuned SR where signal sidebands appear in phase and amplitude quadratures and in any linear combination of both.

Fig.~\ref{sqzinput}\,b compares detuned SR with detuned TSR technique in combination with the injection of squeezed fields, as shown in Fig.~\ref{schema}. Here we assume a broadband squeezed field with squeezing parameter $r\!=\!1$, which corresponds to about 8\,dB squeezing of noise variance.
Again for frequencies above 40\,Hz, TSR achieves a lower noise spectral density than detuned SR. Comparing TSR with and without squeezed states input shows that the TSR technique provides quantum noise reduction from squeezed states over a wide band without additional filter cavity. This advantage is a direct result of the resonance doublet symmetrically arranged around the carrier frequency.
Note that for all curves the squeezed states are not optimized for a
simultaneous reduction of radiation pressure noise. i.e. a reduction of quantum noise in the frequency band below 40\,Hz. Such an optimization is possible for the detuned SR as well as for the detuned TSR case, and requires additional filter cavities in both cases.  However, this regime is also strongly affected by thermal noise~\cite{BGV99}, electronic noise from control loops and seismic noise. It might therefore turn out that the slightly increased noise spectral density at frequencies below 40\,Hz in TSR will not be significant.

We now consider the application of the TSR technique to GW detectors. In that case the spectral shape of the quantum noise should be considered with respect to other noise sources. Since all current and planned interferometric GW detectors are designed for room temperature operation, thermal noise gives a significant contribution to the overall noise floor~\cite{BGV99}, and the resonant sideband frequency of the TSR topology should be chosen and optimized with respect to the detector's thermal noise. Thermal noise arises from thermally driven motion of the suspended mirror test masses and typically falls off towards higher frequencies. Since the rate of detectable gravitational wave signals are expected to decrease with increasing frequency, one may design the resonance peak, having a fixed frequency, fitting the falling thermal noise. If nevertheless one wishes to tune the frequency of the resonance peak, as is possible in the detuned SR technique, mirror TSRM needs to be replaced by a short cavity, for example in the form of an etalon. The reflectivity of such an etalon and therefore the resonance frequency of the TSR scheme, can be changed in situ, for example by changing the etalon's temperature.

In conclusion we propose a new interferometer technique aiming at optimum quantum noise reduction at
intermediate and upper (shot noise limited) detection frequencies of ground based gravitational wave
detectors. The new technique requires a single additional mirror on top of the current GEO\,600 topology and
solves two problems at the same time. With this technique upper and lower signal sidebands are resonantly
enhanced simultaneously and squeezed states can be used without setting up another long-baseline filter
cavity.


\begin{thebibliography}{12}
\bibitem{ligo} D.~Sigg (for the LIGO Science Collaboration), Class. Quantum Grav. {\bf 23}  S51 (2006) 

\bibitem{tama} M.~Ando \emph{et. al.}, Phys.~Rev.~Lett. {\bf 86}, 3950 (2001) 

\bibitem{virgo} F.~Acernese \emph{et al.}, Class. Quantum Grav. {\bf 23} S635 (2006) 

\bibitem{geo} S.~Hild  (for the LIGO Scientific Collaboration), Class. Quantum Grav. {\bf 23} S643 (2006) 

\bibitem{HSMSWWSRD98}G.~Heinzel, K.~A~Strain, J.~Mizuno, K.~D.~Skeldon, B.~Willke, W.~Winkler, R.~Schilling, A.~R{\"u}diger, and K.~Danzmann, \prl {\bf 81}, 5493 (1998)

\bibitem{advligo} A.~Weinstein, Class. Quantum Grav. {\bf 19} 1575-1584 (2002) 

\bibitem{Cav81} C.~M.~Caves, Phys. Rev. D {\bf 23}, 1693 (1981) 

\bibitem{VCHFDS06} H.~Vahlbruch, S.~Chelkowski, B.~Hage, A.~Franzen, K. Danzmann, and R. Schnabel, Phys. Rev. Lett. {\bf 97}, 011101 (2006) 

\bibitem{HCCFVDS03}J.~Harms, Y.~Chen, S.~Chelkowski, A.~Franzen, H.~Vahlbruch, K.~Danzmann, and R.~Schnabel, Phys. Rev. D {\bf 68}, 042001 (2003) 

\bibitem{KLMTV01} H.~J.~Kimble, Y.~Levin,  A.~B.~Matsko, K.~S.~Thorne, and S.~P.~Vyatchanin, Phys. Rev. D {\bf 65}, 022002 (2001)

\bibitem{SHSD04} R.~Schnabel, J.~Harms,  K.~A.~Strain, and K.~Danzmann, Class. Quantum Grav. \textbf{21}, S1045 (2004)

\bibitem{MAD} This feature has previously been noticed by B.~J.~Meers and R.~W.~P.~Drever in unpublished work.

\bibitem{VCHFDS05} H.~Vahlbruch, S.~Chelkowski, B.~Hage, A.~Franzen, K.~Danzmann, and R. Schnabel, Phys. Rev. Lett. {\bf 95}, 211102 (2005)

\bibitem{VinePRL03} G.~de~Vine, M.~Gray, D.~E.~McClelland, Y.~Chen, and S.~Whitcomb, Phys. Lett. A {\bf 316}, 17 (2003)

\bibitem{ThuringPRE72} A.~Th{\"u}ring, H.~L{\"u}ck, and K.~Danzmann, Phys. Rev. E {\bf 72}, 066615 (2005) 

\bibitem{BuonannoPRD65} A.~Buonanno and Y.~Chen, Phys. Rev. D {\bf 65}, 042001 (2002)

\bibitem{BGV99} V.~B.~Braginsky, M.~L.~Gorodetsky, and S.~P.~Vyatchanin, Phys. Lett. A {\bf 264}, 1 (1999)


\end{thebibliography}
\end{document}